%% file: Filament_paper.tex
\documentclass[useAMS,usenatbib]{mn2e}
\usepackage{amssymb,amsmath,epsfig,times,natbib}

\title[X-ray Analysis of Filaments]{X-ray Analysis of Filaments in Galaxy Clusters}
\author[S. A. Walker et al.]{S. A. Walker$^1$\thanks{Email: 
    swalker@ast.cam.ac.uk}, P. Kosec$^1$,  A. C. Fabian and  J. S. Sanders$^2$\\
  $^1$Institute of Astronomy, Madingley Road, Cambridge CB3 0HA \\
  $^2$Max-Planck-Institute fur extraterrestrische Physik, 85748 Garching, Germany \\
      \\
   \\
   \\
}
\date{}

\begin{document}

\maketitle

\begin{abstract}
We perform a detailed X-ray study of the filaments surrounding the brightest cluster galaxies in a sample of nearby galaxy clusters using deep Chandra observations, namely the Perseus, Centaurus and Virgo clusters, and Abell 1795. We compare the X-ray properties and spectra of the filaments in all of these systems, and find that their Chandra X-ray spectra are all broadly consistent with an absorbed two temperature thermal model, with temperature components at 0.75 and 1.7 keV. We find that it is also possible to model the Chandra ACIS filament spectra with a charge exchange model provided a thermal component is also present, and the abundance of oxygen is suppressed relative to the abundance of Fe. In this model, charge exchange provides the dominant contribution to the spectrum in the 0.5-1.0 keV band. However, when we study the high spectral resolution RGS spectrum of the filamentary plume seen in X-rays in Centaurus, the opposite appears to be the case. The properties of the filaments in our sample of clusters are also compared to the X-ray tails of galaxies in the Coma cluster and Abell 3627. In the Perseus cluster, we search for signs of absorption by a prominent region of molecular gas in the filamentary structure around NGC 1275. We do find a decrement in the X-ray spectrum below 2 keV, indicative of absorption. However the spectral shape is inconsistent with this decrement being caused by simply adding an additional absorbing component. We find that the spectrum can be well fit (with physically sensible parameters) with a model that includes both absorption by molecular gas and X-ray emission from the filament, which partially counteracts the absorption.
\end{abstract}


\section{Introduction}

Filaments of material which are bright in their X-ray and H$\alpha$ emission have been found surrounding the brightest cluster galaxies (BCGs) in a number of galaxy clusters. Detailed studies of individual, nearby systems such as NCG 1275 in the Perseus cluster have revealed that the histories of these filaments are intertwined with those of the X-ray cavities in the intracluster medium (ICM) produced from AGN feedback (e.g. \citealt{Fabian2003}, \citealt{Hatch2006}). The filaments are dragged outwards from the centre of the BCG by the buoyantly rising bubbles, before falling back. The morphologies of these filaments can be used as streamlines, allowing us to trace the flow of gas in the ICM. 

The remarkable linearity of the filaments around NGC 1275, some of which are over 6kpc long yet only 70 pc wide, indicates that they are supported by magnetic fields which stabilise them against tidal shear and dissipation into the surrounding, hotter ICM (\citealt{Fabian2008}). 

\begin{figure*}
  \begin{center}
    \leavevmode
    \hbox{
     \epsfig{figure=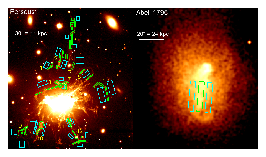,
        width=1.0\linewidth}
           }
       \hbox{
     \epsfig{figure=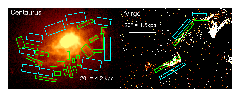,
        width=1.0\linewidth}
        }
      \caption{Images of the filaments (in H$\alpha$ except for A1795, for which the Chandra X-ray data are shown) and the regions used to study them. Green boxes show the `on'-filament regions, while the cyan boxes for the local background (`off'-filament) regions used for the X-ray analysis of each filament. The clusters shown are Perseus (top left), Abell 1795 (top right), Centaurus (bottom left) and Virgo (bottom right). }
        \label{filaments_regions_images}
  \end{center}
\end{figure*}

The excitation mechanism of the filaments remains uncertain. Around NGC 1275 in the Perseus cluster, photoionisation is an unlikely mechanism as the sources of ionizing radiation are not sufficiently luminous, and there are no obvious clusters of stars associated with the filaments. The filaments also have an emission spectrum which is different from any known Galactic nebula. Around other BCGs, some filaments are seen to be associated with ongoing star formation (for example, \citealt{McNamara2004}, \citealt{ODea2004}, \citealt{ODea2010}, \citealt{McDonald2011}, \citealt{Tremblay2015} and \citealt{Donahue2015}).

\citet{Johnstone2007} studied the H$_{2}$ emission from the filaments around NGC 1275 using \emph{Spitzer} and ground-based observations and found that the level excitation energy was correlated with the derived excitation temperature. It was then found that this correlation could be accounted for by collisional heating from ionizing particles (\citealt{Ferland2008}, \citealt{Ferland2009}), with agreement within a factor of 2 over a wide range of line ratios. It is most likely that the optical/infrared/radio emission-line spectrum is produced by gas exposed to ionizing particles.

\citet{Fabian2011_filaments} studied a prominent filament to the north of NGC 1275 and found that the surface radiative flux from the filament is comparable to the energy flux on it from the surrounding hot gas. Given this remarkable similarity, \citet{Fabian2011_filaments} argued that the most likely energy source of these ionizing particles is from the hot gas surrounding the filaments. \citet{Fabian2011_filaments} proposed that the hot gas surrounding the cold filaments is able to penetrate into them through the process of reconnection diffusion (\citealt{Lazarian2010}, \citealt{Lazarian2011}), which provides a mechanism through which the hot gas can penetrate into the cold filaments despite the magnetic fields which support the filaments. This penetration process provides the secondary electrons within the cold filaments which excite the observed submillimetre radiation though UV emission. The interpenetration of the cold gas in the filaments by the surrounding hot gas provides another mechanism through which the hot gas can lose energy, which may mean that gas can cool from the hot ICM at a rate higher than that deduced from just X-ray spectroscopy.

In this mechanism, the hot ionised gas surrounding the filaments penetrates and mixes with the cold neutral gas in the filaments. Charge exchange will then dominate, with electrons being transferred from the cold neutral gas into excited levels of the impinging hot ions. X-ray line emission then results from the cascade of the electrons down to lower energy levels. This process has been observed in other astrophysical environments, such as for example the X-ray emission seen from comets caused by their hydrogen haloes interacting with highly charged ions in the solar wind (e.g. \citealt{Dennerl1997}). 

\citet{Fabian2011_filaments} found that X-ray spectrum of the northern filament of NGC 1275 was dominated by line emission, and could be well fitted with a two temperature model with components at $\sim$0.7 and $\sim$1.5 keV. In the proposed reconnection diffusion model, charge exchange is expected to contribute to the soft X-ray emission of the filaments through Fe-L and O emission lines. However, \citet{Fabian2011_filaments} found that the predicted soft X-ray flux is around a factor of 30 lower than that which is actually observed, if it is assumed that each incoming ion has only one chance of charge exchange per stage. One possibility is that the cold filaments are full of tiny strands, which could increase the rate of interaction resulting in charge exchange emission.

Here we explore the X-ray spectra of the filaments around the BCGs of nearby galaxy clusters for which very deep, high spatial resolution Chandra observations are available. These are namely the Perseus cluster, the Centaurus cluster, the Virgo cluster and Abell 1795. We investigate whether or not the X-ray spectra of the filaments in these different systems are similar, to try understand and constrain the underlying mechanism responsible for their soft X-ray emission. We test whether the latest charge exchange model ACX (AtomDB Charge Exchange model, \citealt{ACXpaper}) is able to to model the X-ray spectra of the filaments in Perseus, and the RGS spectrum of the filamentary plume in the Centaurus cluster.

We use a standard $\Lambda$CDM cosmology with $H_{0}=70$  km s$^{-1}$
Mpc$^{-1}$, $\Omega_{M}=0.3$, $\Omega_{\Lambda}$=0.7. All errors unless
otherwise stated are at the 1 $\sigma$ level. 

\section{Data Reduction}

We stacked all of the available Chandra archive data on the Perseus cluster (890 ks), the Centaurus cluster (760 ks), the Virgo cluster (500 ks) and Abell 1795 (700 ks). The observations used are tabulated in table \ref{All_obsids} in appendix \ref{sec:appendix}, and these were reduced using CIAO version 4.7. The events files were reprocessed using the task CHANDRA\_REPRO. Light curves in the 0.5-7.0 keV band were examined, and the routine LC\_SIGMA\_CLIP was used to remove any periods of flaring where the count rate differed from the mean by more than 2 $\sigma$. The resulting clean exposure times are tabulated in table \ref{All_obsids}.

\begin{table}
\begin{center}
\caption{Coordinates, redshift and absorption data of objects used in analysis.}
\label{Object_Information}
\leavevmode
\begin{tabular}{ l|l | l | l | l} \hline \hline
Object&RA&Dec&z&N$_{\rm H}$/(cm$^{-2}$) \\ \hline
Perseus&03h19m47.2s&+41d30m47s&0.017900&1.4$\times 10^{21}$\\ 
Centaurus&12h48m47.9s&-41d18m28s&0.011400&8.30$\times 10^{20}$\\
M87&12h30m49.4s&+12d23m28s&0.004283&2.07$\times 10^{20}$\\
A1795&13h48m53.0s&+26d35m44s&0.062476&1.21$\times 10^{20}$\\
Coma&12h59m48.7s&+27d58m50s&0.023100&8.43$\times 10^{19}$\\
A3627&16h14m22.5s&-60d52m07s&0.015700&1.81$\times 10^{21}$\\\hline
\end{tabular}
\end{center}
\end{table}

\section{Methodology}

\begin{figure}
  \begin{center}
    \leavevmode
     \vbox{ \epsfig{figure=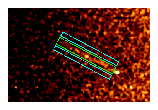,
        width=1.0\linewidth}
        \epsfig{figure=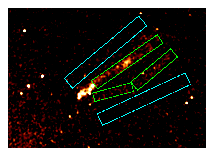,
        width=1.0\linewidth}
              \epsfig{figure=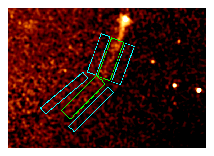,
        width=1.0\linewidth}
                }
      \caption{X-ray images of the galaxy tails studies, along with the spectral extraction regions for the tails (green) and the local background regions used (cyan). The top panel shows the galaxy tail in the Coma cluster, while the middle and bottom panels show galaxy tails in Abell 3627.}
      \label{galaxy_tail_regions}
  \end{center}
\end{figure}

The filaments around the BCGs in the systems we study are most easily seen through their H$\alpha$ emission. We therefore choose the regions to study based on these H$\alpha$ images (except for A1795, for which the X-ray image is used), defining the `on' and `off' filament regions shown in the panels in Fig \ref{filaments_regions_images}. These panels show Perseus (top left, H$\alpha$ WIYN telescope image from \citealt{Conselice2001}), Abell 1795 (top right, Chandra 0.5-1.5 keV image), Centaurus (bottom left, ESO Multi-Mode Instrument H$\alpha$ image from \citealt{Crawford2005}) and Virgo (bottom right, H$\alpha$ HST WFPC2 image), which were used for extracting X-ray spectra from the Chandra data. The `off'-filament regions, shown in cyan were used to provide a local background and were subtracted from the `on'-filament regions, scaling for the difference in extraction size area between the foreground and the background. This allows the X-ray spectra of the excess emission from the filaments to be explored. For the H$\alpha$ image of Virgo, the galaxy emission from the BCG has been subtracted to emphasise the filaments. 

We also explore three X-ray galaxy tails seen in the Coma cluster and Abell 3627, to see how their spectra compare to the filament spectra. The regions used in the spectral analysis of these tails are shown in the images in Fig. \ref{galaxy_tail_regions}.

For each spectral extraction region, the X-ray spectra were extracted from each individual Chandra observation, together with the appropriate ARF and RMF calibration files for each observation using the tools MKWARF and MKACISRMF respectively. The spectra from all of the observations of a particular region were then combined and weighted appropriately using the ARFs to produce the stacked spectrum for each region, while the individual ARFs and RMFs for each region were also combined with the correct weighting. Due to differences in the effective areas of the ACIS-S and ACIS-I detectors, the data from these two detectors were not added together. Instead, when both ACIS-S and ACIS-I data were used for the same target, we found a total ACIS-S spectrum and a total ACIS-I spectrum, and these two were fitted simultaneously in XSPEC.

We fitted each local background subtracted filament region with an absorbed two component thermal model, \textsc{phabs(apec+apec)}, as has been employed in \citet{Fabian2011_filaments}. The metal abundances of the two components were tied together, allowing the temperatures and normalisations of each APEC component to be free. The column density was fixed to the LAB survey value (\citealt{LABsurvey}) for each cluster studied, and these are tabulated in table \ref{Object_Information}, along with the values to which the redshift was fixed when fitting.

\subsection{Perseus}

\begin{figure}
  \begin{center}
    \leavevmode
\epsfig{figure=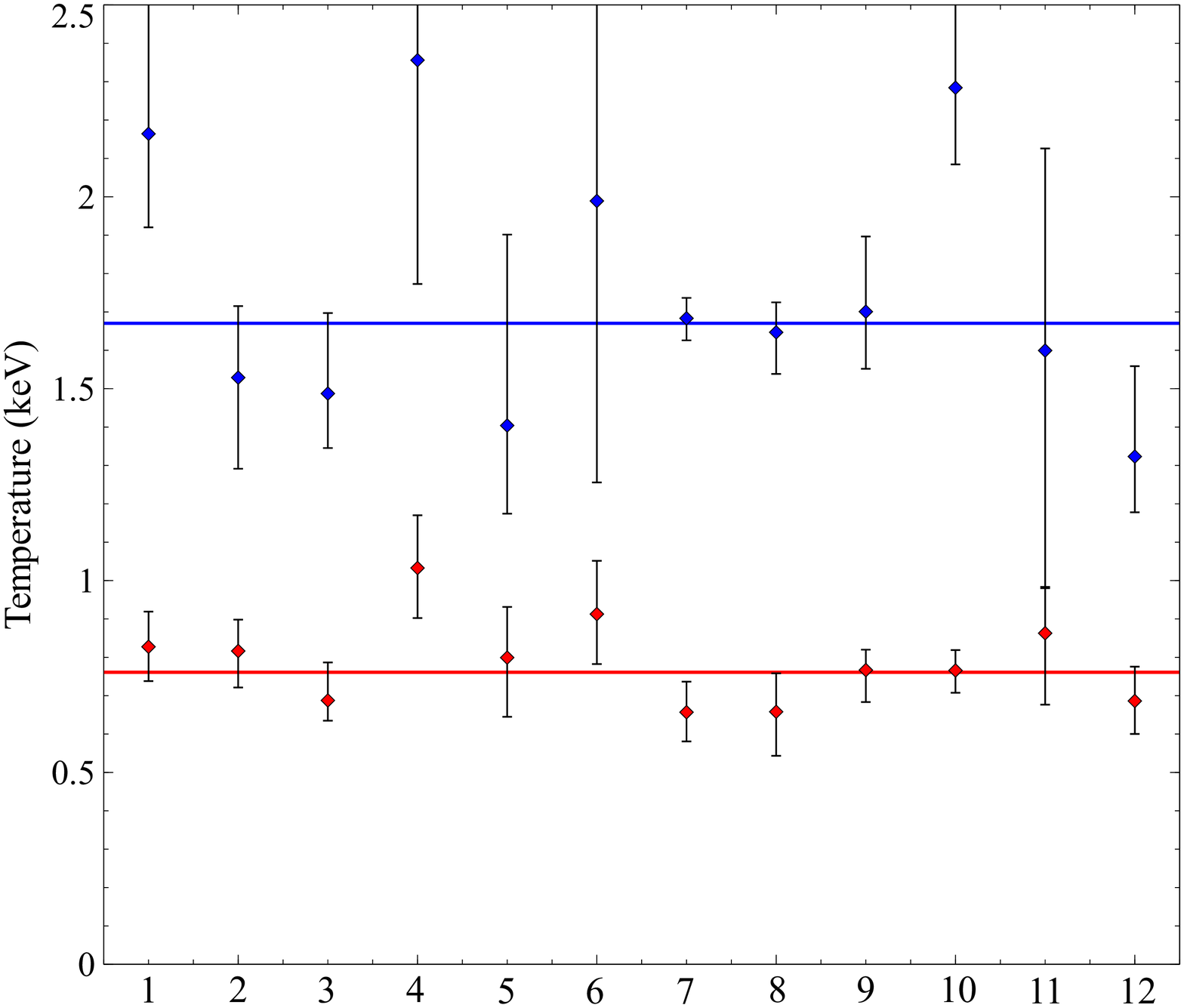,
        width=1.0\linewidth}
      \caption{Plot of temperatures of filaments in the Perseus cluster. Blue points mark the temperature of the hotter component, red points represent the lower temperature component.}
        \label{temps_Per}
  \end{center}
\end{figure}

For the Perseus cluster, the Chandra data are sufficiently deep and the filament emission is sufficiently bright that we can study each filament individually. As shown in the top left hand panel of Fig. \ref{filaments_regions_images}, we identify 12 filaments in Perseus which we study individually. 

The best fitting parameters of the two thermal component model for the 12 filaments in Perseus are tabulated in table \ref{Perseus_results}, and the best fitting temperatures are plotted in Fig. \ref{temps_Per}. In all of the systems the two temperature model provides a good fit. The lower temperature components of the 12 filaments are all in good agreement, and are all reasonably consistent with the average temperature of 0.77 keV (shown by the red line in Fig. \ref{temps_Per}). The spread of the higher temperature component is higher, but again for all of the filaments these are in reasonable agreement with the average upper temperature of 1.78 keV (shown by the blue line in Fig. \ref{temps_Per}). 

\begin{table*}
\begin{center}
\caption{Fitting parameters for all filaments in the Perseus cluster.}
\label{Perseus_results}
\leavevmode
\begin{tabular}{ l|l | l | l | l | l | l} \hline \hline
Number&Temperature 1 (keV)&Temperature 2 (keV)&Abundance (solar)&Norm 1&Norm 2&Norm ratio\\ \hline
1&$0.83^{+0.09}_{-0.09}$&$2.16^{+0.35}_{-0.25}$&$0.43^{+0.24}_{-0.17}$&$4.66^{+2.69}_{-1.56}\times10^{-6}$&$2.02^{+0.34}_{-0.32}\times10^{-5}$&$4.33^{+2.60}_{-1.60}$\\
2&$0.82^{+0.08}_{-0.10}$&$1.53^{+0.19}_{-0.24}$&$1.54^{+3.20}_{-0.78}$&$4.13^{+3.62}_{-2.69}\times10^{-6}$&$8.01^{+4.91}_{-4.42}\times10^{-6}$&$1.94^{+2.08}_{-1.65}$\\
3&$0.69^{+0.10}_{-0.05}$&$1.49^{+0.21}_{-0.14}$&$1.00^{+2.44}_{-0.54}$&$6.19^{+5.93}_{-1.44}\times10^{-6}$&$1.99^{+1.23}_{-0.70}\times10^{-5}$&$3.21^{+3.66}_{-1.35}$\\
4&$1.03^{+0.14}_{-0.13}$&$2.36^{+1.22}_{-0.59}$&$0.13^{+0.15}_{-0.07}$&$1.20^{+1.01}_{-0.60}\times10^{-5}$&$1.19^{+0.46}_{-0.64}\times10^{-5}$&$0.99^{+0.92}_{-0.72}$\\
5&$0.80^{+0.13}_{-0.16}$&$1.40^{+0.50}_{-0.23}$&$2.35^{+2.30}_{-1.68}$&$1.13^{+1.19}_{-0.84}\times10^{-6}$&$2.04^{+4.27}_{-0.80}\times10^{-6}$&$1.81^{+4.23}_{-1.51}$\\
6&$0.91^{+0.14}_{-0.13}$&$1.99^{+5.51}_{-0.74}$&$0.16^{+0.34}_{-0.10}$&$1.13^{+1.17}_{-0.69}\times10^{-5}$&$1.08^{+0.62}_{-0.95}\times10^{-5}$&$0.96^{+1.13}_{-1.02}$\\
7&$0.66^{+0.08}_{-0.08}$&$1.68^{+0.06}_{-0.06}$&$0.69^{+0.18}_{-0.15}$&$1.07^{+0.21}_{-0.17}\times10^{-5}$&$1.01^{+0.14}_{-0.13}\times10^{-4}$&$9.44^{+2.20}_{-1.93}$\\
8&$0.66^{+0.10}_{-0.12}$&$1.65^{+0.08}_{-0.11}$&$0.42^{+0.16}_{-0.13}$&$5.23^{+1.31}_{-1.02}\times10^{-6}$&$5.01^{+0.82}_{-0.68}\times10^{-5}$&$9.58^{+2.87}_{-2.27}$\\
9&$0.77^{+0.06}_{-0.09}$&$1.70^{+0.20}_{-0.15}$&$0.44^{+0.20}_{-0.14}$&$1.12^{+0.35}_{-0.29}\times10^{-5}$&$3.71^{+0.65}_{-0.54}\times10^{-5}$&$3.31^{+1.17}_{-0.97}$\\
10&$0.77^{+0.06}_{-0.06}$&$2.28^{+0.27}_{-0.20}$&$1.17^{+0.56}_{-0.36}$&$6.57^{+2.28}_{-1.78}\times10^{-6}$&$6.32^{+1.17}_{-1.15}\times10^{-5}$&$9.62^{+3.78}_{-3.14}$\\
11&$0.86^{+0.12}_{-0.19}$&$1.60^{+0.53}_{-0.62}$&$2.18^{+2.18}_{-1.37}$&$9.75^{+7.15}_{-4.87}\times10^{-7}$&$1.61^{+4.39}_{-1.03}\times10^{-6}$&$1.65^{+4.66}_{-1.34}$\\
12&$0.69^{+0.09}_{-0.09}$&$1.32^{+0.24}_{-0.15}$&$5.00^{+5.00}_{-3.78}$&$1.14^{+1.22}_{-0.22}\times10^{-6}$&$1.77^{+4.64}_{-0.52}\times10^{-6}$&$1.55^{+4.40}_{-0.54}$\\ \hline
\end{tabular}
\end{center}
\end{table*}

\begin{figure}
 \begin{center}
    \leavevmode
    \vbox{
      \epsfig{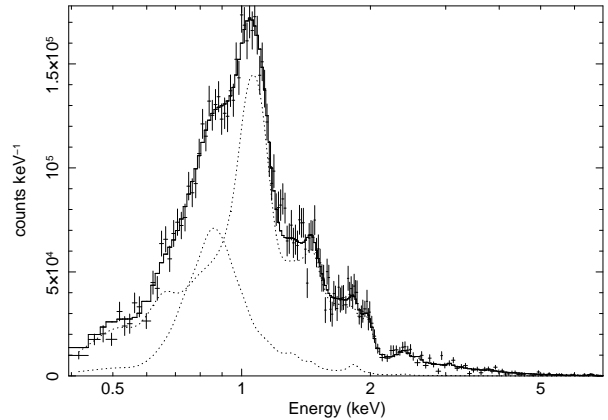}
        }
      \caption{ The spectrum of all filaments (stacked) in Perseus that have been fitted with a two-temperature apec model. }
      \label{Filaments_stack_Perseus}
  \end{center}
\end{figure}

Since the individual filaments appear to have similar spectral shapes, we stack them all together to produce a spectrum with the highest possible signal to noise, and this is shown in Fig. \ref{Filaments_stack_Perseus}, showing components of the best fitting two component thermal model.

\subsection{Virgo, Centaurus and Abell 1795 filaments}

For the other clusters we study, the Chandra data are not sufficiently deep to allow individual filaments to be studied. Instead we obtain a stacked X-ray spectrum of all of the filaments in each cluster, and fit these with the same absorbed two thermal component model as before. In each case the model provides a good fit to the data, and the best fitting temperature components are plotted in Fig. \ref{temps_all}. In each case the lower temperature components appear reasonable consistent with one another. There is more scatter in the higher temperature component, but again these are in broad agreement.

\subsection{Galaxy tails in the Coma cluster and Abell 3627}

\begin{figure}
  \begin{center}
    \leavevmode
      \epsfig{figure=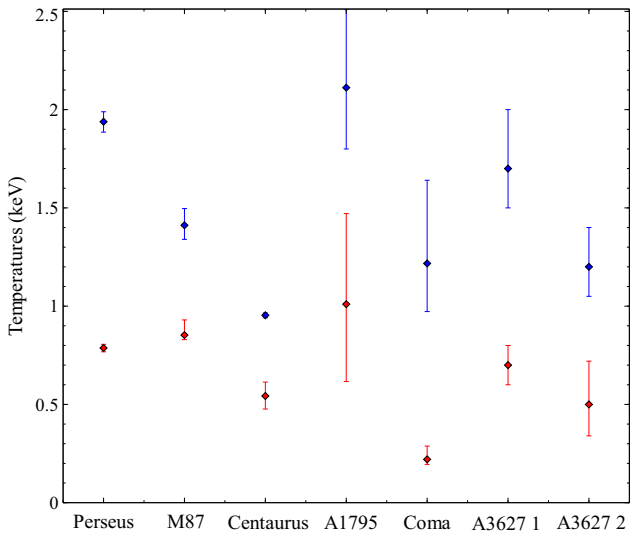,
        width=1.0\linewidth}
      \caption{Plot of temperatures of the best fit two temperature models of filaments in Perseus, M87, Centaurus and Abell 1795, and of the galaxy tails in Coma and Abell 3627. Blue points mark the temperature of the hotter component, red points represent the lower temperature component.}
      \label{temps_all}
  \end{center}
\end{figure}

The spectra of galaxy tails in the Coma cluster (originally presented in \citealt{Sanders2014}) and Abell 3627 (originally presented in \citealt{SunA3627} and \citealt{ZhangA3627}) were extracted as well, for comparison with the filaments in other clusters. 
The tail in the Coma cluster (Fig. \ref{galaxy_tail_regions} top panel), can be fitted with a two-temperature model, with a colder component of temperature $0.22^{+0.07}_{-0.03}$ keV and hotter component of $1.22^{+0.42}_{-0.25}$ keV. The abundance is $0.7^{+2.0}_{-0.0.5}$ Z$_{\odot}$. 
For the first tail in Abell 3627 (Fig. \ref{galaxy_tail_regions}, middle panel), we find that the best fit two temperature model is consistent with the findings of \citet{SunA3627}, with components at 0.7$^{+0.1}_{-0.1}$ keV and 1.7$^{+0.3}_{-0.2}$ keV and a best fit metal abundance of 0.2$^{+0.3}_{-0.1}$Z$_{\odot}$.
The two temperature fit to the second tail in Abell 3627 is consistent with the findings of \citet{ZhangA3627}, with temperature components at $0.5^{+0.22}_{-0.16}$ keV and $1.2^{+0.2}_{-0.15}$ keV and an abundance of 0.3$^{+0.2}_{-0.1}$ Z$_{\odot}$. These temperature are also plotted and compared in Fig. \ref{temps_all}.

\section{Comparison to charge exchange models}

\subsection{Comparing Chandra ACIS spectra of Perseus filaments}
\label{Perseus_fil_CX}
\begin{figure*}
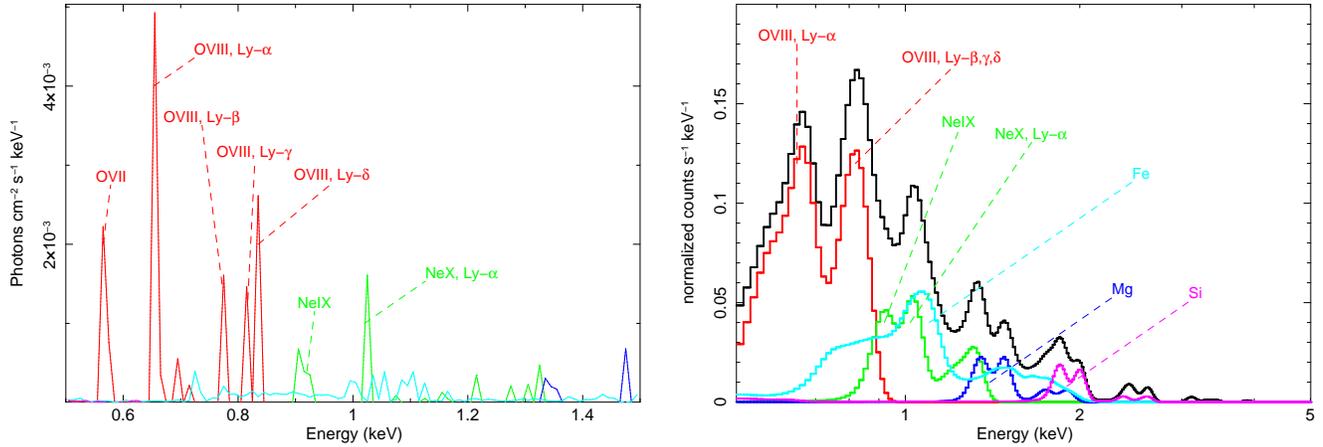

 \begin{center}
    \leavevmode
    \hbox{
      \epsfig{figure=Fig6a.eps,width=0.33\linewidth,angle=-90}
            \epsfig{figure=Fig6b.eps,width=0.33\linewidth,angle=-90}
        }
      \caption{\emph{Left}:The ACX spectral model, with the main lines in the energy range of interest labelled. \emph{Right}: ACX model folded through the Chandra response, with the main charge exchange lines labelled. In this model the relative abundances of all of the metals are set equal to one another. }
      \label{ACX_model}
  \end{center}
\end{figure*}

Here we test whether the X-ray spectra of the filaments can be modelled with the charge exchange model ACX\footnote{http://www.atomdb.org/CX/} (\citealt{ACXpaper}). The ACX model in the energy range of interest is shown in the left hand panel of Fig. \ref{ACX_model}, with the main emission lines labelled. The result of folding this model through the Chandra response is shown in the right hand panel of Fig. \ref{ACX_model}. In this charge exchange model we have set the equilibrium ion population temperature to the temperature of the hot ICM surrounding the filaments, which is 4 keV.

\begin{figure*}
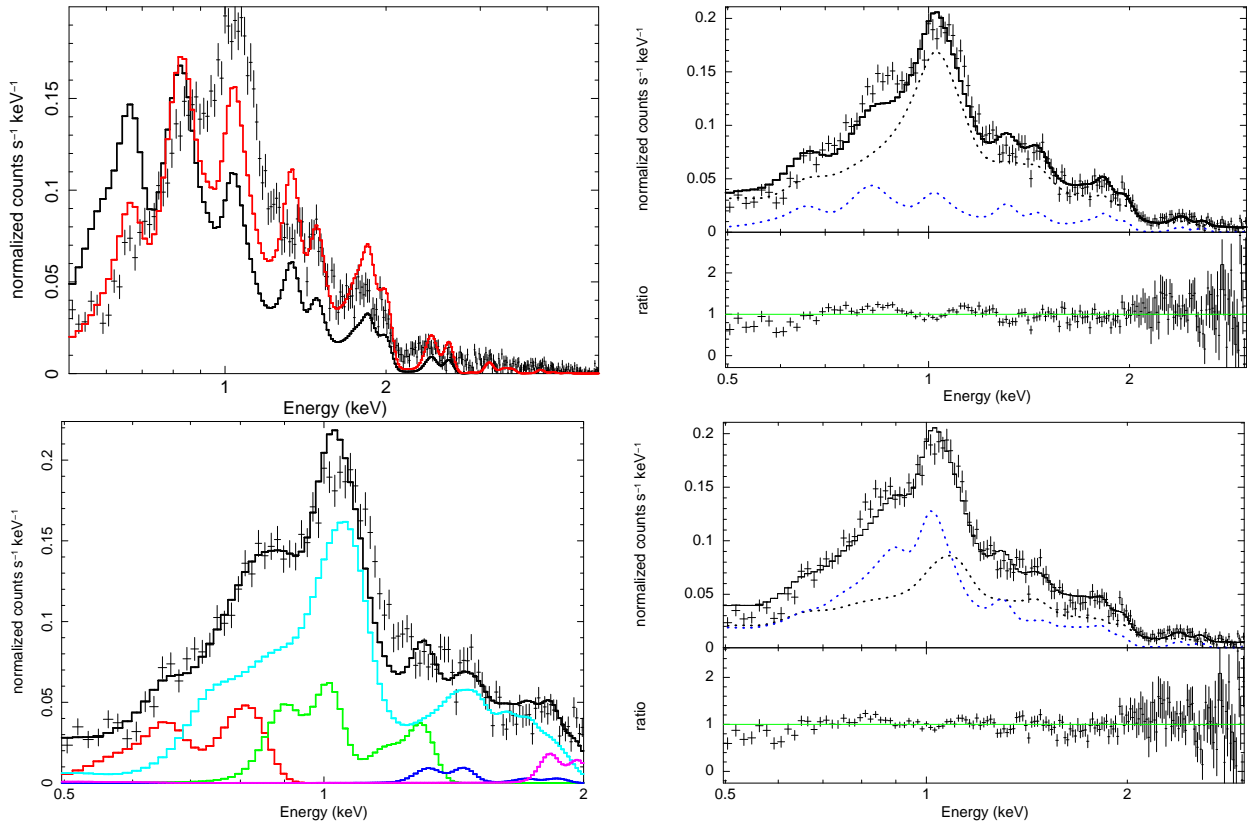

 \begin{center}
    \leavevmode
    \hbox{
     \epsfig{figure=Fig7a.eps,width=0.31\linewidth,angle=-90}
      \epsfig{figure=Fig7b.eps,width=0.3\linewidth,angle=-90}
        }
            \hbox{
\epsfig{figure=Fig7c.eps,width=0.3\linewidth,angle=-90}
     \epsfig{figure=Fig7d.eps,width=0.3\linewidth,angle=-90}

         }
      \caption{ \emph{Top left}: Same spectrum as in Fig. \ref{Filaments_stack_Perseus}, but with a charge exchange model overplotted with different levels of absorption, allowing us to compare the spectral shape. The solid black line is for a model with the LAB survey level of absorption (0.14$\times10^{22}$ cm$^{-2}$). Letting the absorbing column be free and fitting results in the solid red line, for which the column is 0.35$\times10^{22}$ cm$^{-2}$. \emph{Top right}: Fitting the same spectrum but now adding an absorbed thermal component (dashed black curve) to the charge exchange model (dashed blue curve). The fit remains very poor. \emph{Bottom left}: Fitting the filament spectrum with an absorbed variable abundance charge exchange model. The colour scheme of the lines is the same as in Fig. \ref{ACX_model}. \emph{Bottom right}: Fitting with an absorbed variable abundance charge exchange model (blue dashed curve) plus an absorbed apec model (black dashed curve). }
      \label{Filaments_stack_Perseus_CX}
  \end{center}
\end{figure*}

In Fig. \ref{Filaments_stack_Perseus_CX} we compare the stacked spectrum of all of the filaments in Perseus to the shape of the charge exchange model ACX. To begin with we used the LAB survey column density of 0.14$\times10^{22}$ cm$^{-2}$, for which the spectrum is shown as the solid black line in the top left hand panel of Fig. \ref{Filaments_stack_Perseus_CX}.  We see that the shape of the charge exchange spectrum below 1.0 keV disagrees significantly with the observed X-ray spectrum of the stacked filaments, due to the OVII and OVIII lines being far brighter in the charge exchange model. 

We then allow the column density to be a free parameter, to see if this can suppress the OVII and OVIII lines sufficiently to produce an adequate fit to the observed spectrum. The resulting attempted fit is shown as the solid red line in the top left hand panel of Fig. \ref{Filaments_stack_Perseus_CX}, for which the column density is 0.36$\times10^{22}$ cm$^{-2}$. We see that the model remains a very poor fit to the data.

Next we add an absorbed APEC component to the model, so that the total model is PHABS(APEC+ACX), to see if we can obtain an acceptable fit. The resulting fit is shown in the right hand panel of Fig. \ref{Filaments_stack_Perseus_CX}, where we have again allowed the absorption of the charge exchange model (dashed blue curve) to be a free parameter. The fit improves on the addition of an APEC component but remains very poor, as can be seen from the residuals, with a reduced $\chi^{2}$ of 2.1 for 166 degrees of freedom in the 0.5-3.0 keV band. In this fit the column density is 0.2$^{+0.01}_{-0.01}\times10^{22}$ cm$^{-2}$, while the APEC component, (which is the dominant component) has kT=1.4 keV and an abundance of 0.2 Z$_{\odot}$.

Next we allow the relative abundances of O, Ne, Mg, Si and Fe to be free parameters, and try to fit the filament spectrum using an absorbed variable abundance charge exchange model, PHABS(VACX). The best fit is shown in the bottom left panel of Fig. \ref{Filaments_stack_Perseus_CX}. The best fit relative abundances for O:Ne:Mg:Si:Fe are
0.25:0.64:0.2:0.4:1.5. The enhancement of Fe relative to O and Ne dramatically improves the fit below 1 keV, and can correctly model the `step' shaped feature between 0.7 and 1.0 keV. The best fit column density is 0.23$^{+0.01}_{-0.01}\times10^{22}$ cm$^{-2}$, and the quality of the fit is only slightly reduced if this is fixed to the LAB survey value of 0.14$^{+0.01}_{-0.01}\times10^{22}$ cm$^{-2}$. However the fit above 1 keV remains poor, most notably the deficiency of the model in the 1-1.3 keV range, leading to a very poor goodness of fit statistic (reduced $\chi^2$=2.98 for 102 degrees of freedom, fitting to the 0.5-2.0 keV band).

To try to improve the fit further, particularly above 1 keV, we add an absorbed APEC component, so that the model becomes PHABS(VACX+APEC). Fitting for the broad 0.5-7.0 keV band, we find that the fit statistic can be improved to a reduced $\chi^{2}$ of 1.26 for 446 degrees of freedom, and the best fit spectrum below 3 keV is shown in the bottom right hand panel of Fig. \ref{Filaments_stack_Perseus_CX}. In this fit the relative abundances of O:Ne:Mg:Si:Fe in the VACX model (blue dashed curve) are 0.2:1.4:0:0.4:2.0, with the oxygen abundance again being suppressed relative to Fe. The column density is fixed to the LAB survey value of 0.14$\times10^{22}$ cm$^{-2}$. The APEC component (black dashed curve) is able to improve the fit dramatically above 1 keV. The temperature of the APEC component is 2.2$^{+0.2}_{-0.2}$ keV, which is reasonable. In this model, the charge exchange component is the dominant component in the 0.5-1.0keV band, contributing 57 percent of the X-ray flux. 

If charge exchange is a significant contributor to the 0.5-1.0 keV band, the constant low temperature, $\sim$0.8 keV, component seen in the two thermal component model fits to the filament spectra (Fig. \ref{temps_Per}) may be modelling line in the charge exchange spectrum originating from the Lyman $\beta$, $\gamma$ and $\delta$ transitions in OVIII. This would provide an explanation for why the same low temperature component is always found in the two temperature thermal fits to the spectra.

We therefore conclude that a model consisting of charge exchange plus an APEC model is a possibility for explaining the Chandra ACIS filament spectrum of the Perseus cluster, with charge exchange being the dominant contributor to the X-ray flux in the 0.5-1.0 keV band. While the fit is not formally as good as for an absorbed two temperature thermal model, further improvements in the charge exchange model may result in an improved fit. Further theoretical work is also needed to determine whether the normalisation of the charge exchange emission can match the observed X-ray flux from the filaments (\citealt{Fabian2011_filaments}).

We note that \citet{Werner2013} have also previously examined the soft X-ray emission from the filaments in the Virgo cluster, and found that they are brightest in the 0.7-0.9 keV band. This is precisely the band in which the ACX charge exchange model predicts there to be significant contributions from the Ly-$\beta$, $\gamma$ and $\delta$ lines of OVIII.

 \begin{figure*}
  \begin{center}
  \hbox{
    \leavevmode
               \epsfig{figure=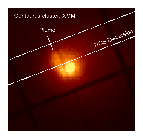, width=0.4\linewidth}
                       \epsfig{figure=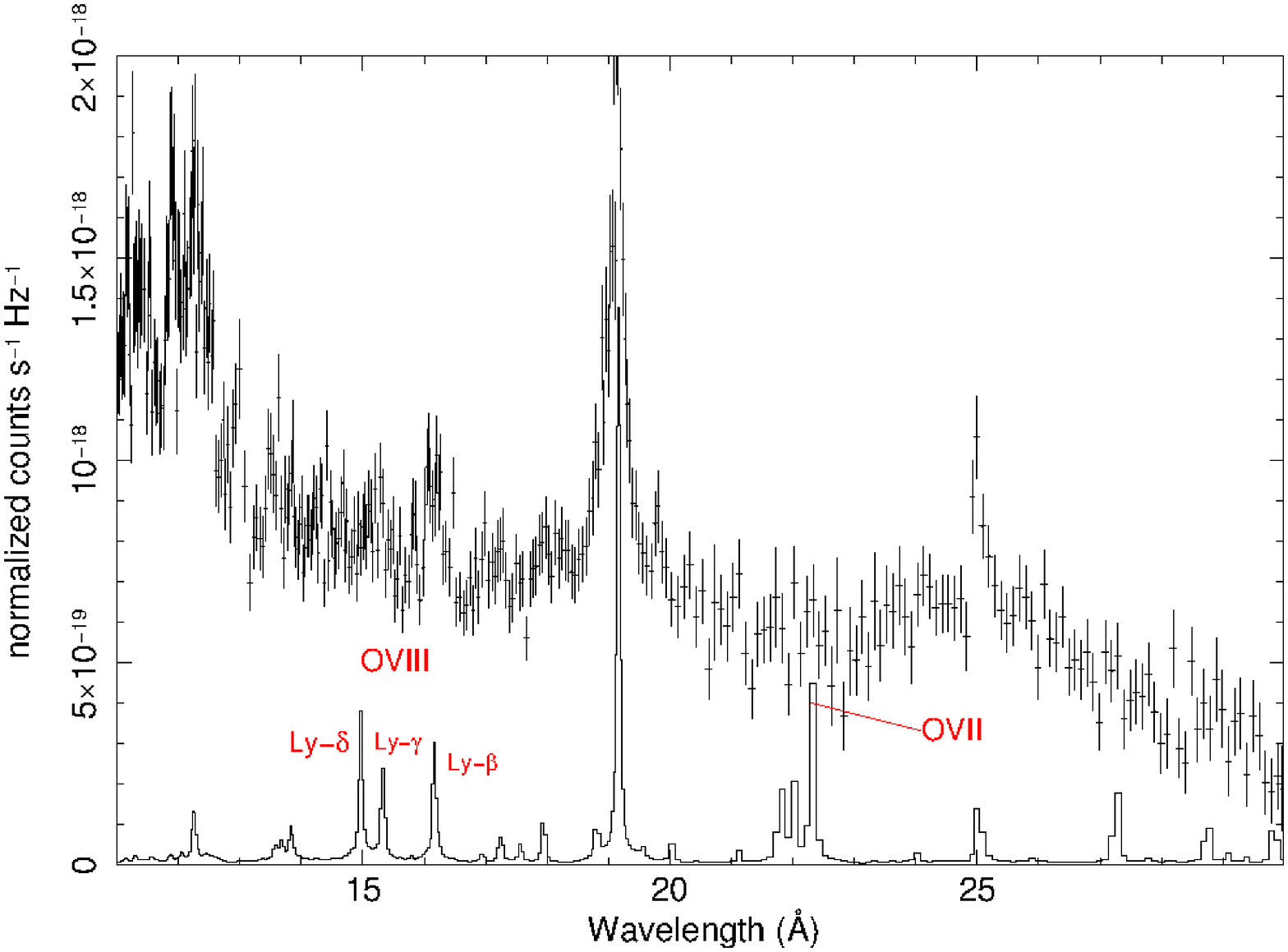, width=0.6\linewidth}

        }      

      \caption{\emph{Left}:XMM-Newton EPIC image of the Centaurus cluster core, showing that the plume direction is perpendicular to the dispersion direction of the RGS. The 90 percent PSF extraction region (a strip of width 60 arcsec) is shown by the dashed white lines. \emph{Right}: RGS spectrum of the plume compared to the emission lines in the ACX charge exchange spectrum. Note that in this figure the model spectrum has not been broadened to account for spatial broadening, and we use it only to show the expected locations of the lines. The spectrum also includes significant continuum emission from the surrounding intracluster medium. We see that the prominent OVII line and the OVIII Ly-$\gamma$ and Ly-$\delta$ lines expected from charge exchange are not present. }
      \label{RGS_Centaurus}
  \end{center}
\end{figure*}

\subsection{Comparing XMM RGS spectra of Centaurus filaments}

Here we compare the RGS spectrum of the X-ray emitting filamentary plume in the Centaurus cluster with the ACX charge exchange model. In the XMM observations we study, (obsids:0046340101 and 0406200101), the roll angle has been set to make the plume direction perpendicular to the RGS dispersion direction (shown in the left hand panel of Fig. \ref{RGS_Centaurus}). This arrangement allows an RGS spectrum for the plume to be obtained which excludes contributions from the nucleus, and this spectrum is shown in the right hand panel of Fig. \ref{RGS_Centaurus}. The data reduction of the RGS data followed the methods described in \citet{Sanders2008}. The coordinates of the extraction position for the spectrum are RA,DEC= (192.2194,-41.303982), and we used an inner 90 per cent PSF extraction, corresponding to a 60 arcsec wide strip across the cluster.  

In the right hand panel of Fig. \ref{RGS_Centaurus} we compare the RGS spectrum of the plume with the ACX charge exchange spectrum. The plume spectrum does not appear to contain obvious evidence of emission from the OVII line which would be expected to be prominent. The Ly-$\beta$, Ly-$\gamma$ and Ly-$\delta$ lines from OVIII are also not clearly present. This suggests that charge exchange is not the dominant contributor to the X-ray emission from the filamentary plume in the Centaurus cluster. 

We emphasise, however, that this X-ray plume does not contain any H-$\alpha$ filaments, as can be seen by comparing the X-ray image in Fig. \ref{RGS_Centaurus} with the H-$\alpha$ image of Centaurus in the bottom left hand panel of Fig. \ref{filaments_regions_images}. The RGS spectrum for Centaurus we present here is therefore not extracted from the same type of environment as the Chandra spectrum of the Perseus filaments we presented in section \ref{Perseus_fil_CX}. This may provide an explanation for the weakness of the any charge exchange lines in the RGS spectrum. We also stress that the RGS spectrum has a significant continuum contribution from the surrounding intracluster medium, which cannot be subtracted accurately, and which may obscure any emission lines from charge exchange. The effective area of the RGS is also reduced in the 20-24 $\rm \AA$ region compared to the rest of the spectrum due to the malfunction of CCD4 of the RGS2 detector, lowering the signal to noise.

\section{Absorption by cold gas}

\begin{figure}
  \begin{center}
    \leavevmode
      \epsfig{figure=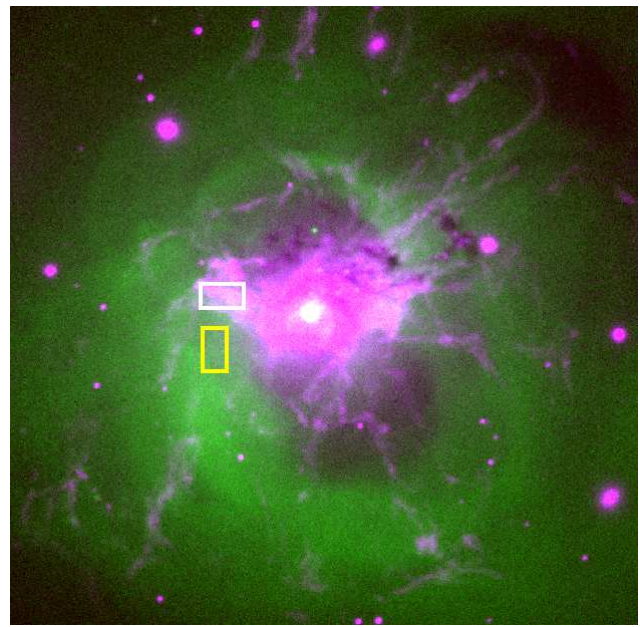,
        width=1.0\linewidth}
      \caption{Composite image of the centre of the Perseus cluster. The H$\alpha$ filaments are shown in pink, while the X-ray emission from Chandra observations is shown in green. The white box is located on a prominent area of molecular gas as studied by \citet{Salome2008}, and we compare the X-ray spectrum from this region with that from a nearby region (yellow box) with no intervening molecular gas along the line of sight. }
      \label{RGB_image_perseus}
  \end{center}
\end{figure}

To explore whether the gas in the filamentary structure be seen in absorption in X-rays, we explore a region in the Perseus cluster, shown by the white box in Fig. \ref{RGB_image_perseus}. This region has been the focus of detailed CO observations (\citealt{Salome2008}), which found an average gas mass in this region of $\sim$5$\times 10^{7}$ M$_{\odot}$. Taking the volume of the molecular gas region to be a cuboid of dimensions 1000pc $\times$ 500pc $\times$ 500 pc, this yields a number density of the molecular gas of $\sim$10 particles cm$^{-3}$. Taking the depth to be the same as its width (500pc), this yields a sizeable expected column density of 10$^{22}$ cm$^{-2}$, which we would expect to be able to discern from X-ray observations. 

\begin{figure}
  \begin{center}
    \leavevmode
      \epsfig{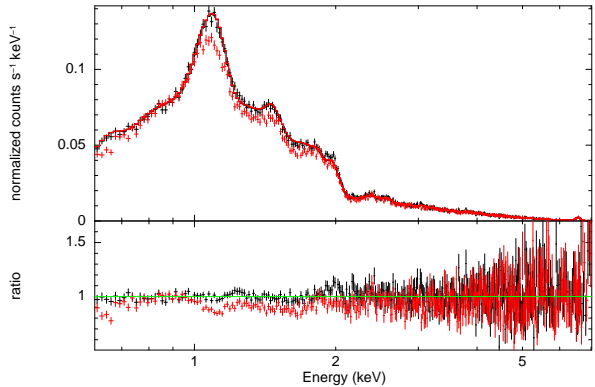}
      \caption{Comparing the X-ray spectra of the molecular gas region (red spectrum, white box in Fig. \ref{RGB_image_perseus}) with that of an offset region (black spectrum, yellow box in Fig. \ref{RGB_image_perseus}). Both spectra have been scaled to account for any differences in the areas in the extraction regions. The best fit model is fit to the black spectrum (for the offset region), and the lower panel shows the residuals to this model. It is clear that in the region with molecular gas (red spectrum) there is a deficit in the X-ray emission at energies below 2 keV.     }
      \label{abs_spec_fig}
  \end{center}
\end{figure}

To test whether this absorption is seen in the X-ray observations, we compared the X-ray spectrum on the filament region (the white box in Fig. \ref{RGB_image_perseus}) with a neighbouring region devoid of filaments (the yellow box in Fig. \ref{RGB_image_perseus}), which also lies outside the X-ray cavities in a region with similar X-ray surface brightness. These spectra are shown in Fig. \ref{abs_spec_fig}, where the red spectrum is from the region with molecular gas we study (the white box in Fig. \ref{RGB_image_perseus}), and the black spectrum is for the nearby offset region (the yellow box in Fig. \ref{RGB_image_perseus}). Both spectra have been scaled to account for differences in their extraction areas. The best fit model shown is fit to the offset region only, and the lower panel of Fig. \ref{abs_spec_fig} shows the ratio to this model. This model is an absorbed two component thermal model, PHABS(APEC1+APEC2), where the column density in the PHABS component is fixed to the LAB survey value of 0.14$\times10^{22}$ cm$^{-2}$. We see that there is clear deficit in the X-ray spectrum of the molecular gas region below 2 keV, and the spectra are essentially identical above 2 keV. 

Qualitatively, this soft X-ray decrement agrees with the idea that absorption by the gas in the filament region is taking place. However the spectral shape of the decrement cannot be modelled accurately using an additional absorption model. As can be seen in the ratio plot in Fig. \ref{abs_spec_fig}, the soft X-ray decrement fluctuates, being at its greatest in the energy range 0.6-0.7 keV and 1-2 keV, but being very small in the energy range 0.8-1.0 keV. This variation in the decrement below 2 keV cannot be fit for by just adding additional absorbing components to the best fit spectrum to the offset region. 

\begin{figure}
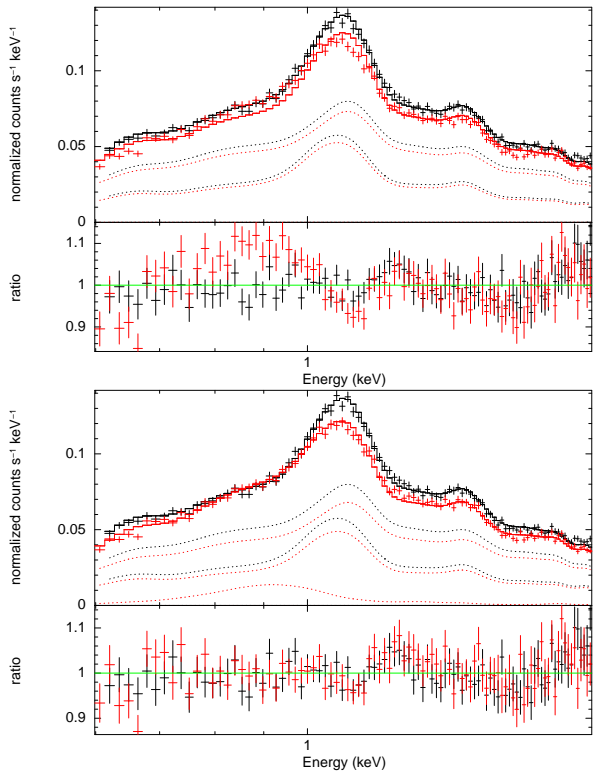

  \begin{center}
    \leavevmode
    
   \vbox{   \epsfig{figure=Fig11a.eps,
        width=0.6\linewidth,angle=-90}
         \epsfig{figure=Fig11b.eps,
        width=0.6\linewidth,angle=-90}
        }
      \caption{The same spectra as shown in Fig. \ref{abs_spec_fig}, but focusing just at the 0.6-2.0 keV soft band. \emph{Top}: Here we attempt to fit the spectrum for the molecular region (red) by just adding a partial covering absorption model to the best fit spectrum to the offset region (the black spectrum). As the residuals show, the resulting fit is very poor. \emph{Bottom}: Same as the top panel but we now add an additional APEC component to represent possible X-ray emission from the filaments (shown as the lowest dashed red line in the spectral model). This results in a good fit.   }
      \label{abs_spec_fig_2}
  \end{center}
\end{figure}

To show that this is the case, we take the best fit model to the offset region (the black spectrum in Fig. \ref{abs_spec_fig}) and attempt to fit it to the molecular gas region just by adding a partial covering absorption component, so that the model becomes PCFABS*PHABS(APEC1+APEC2). All of the original components of the model (the PHABS(APEC1+APEC2) part) are fixed to their values for the offset region. The only free parameters are the column density and covering fraction of the partial covering absorption model, PCFABS.  The result is shown in the top figure of Fig. \ref{abs_spec_fig_2} (the red spectrum). We see that this model is a poor fit to the data.

One possibility is that in addition to absorption, the filamentary material in the region also emits in X-rays, and this acts to partially compensate for the decrement. To test for this possibility, we add into the model an additional soft thermal component to represent possible filament emission. The model is now PCFABS*PHABS(APEC1+APEC2+APEC3), where the PHABS(APEC1+APEC2) is fixed to the best fitting values to the offset spectrum as before. The free parameters are the column density and covering fraction of the partial covering PCFABS model, and the temperature and normalisation of the new thermal component (APEC3) describing possible filament emission. 

This model produces a good fit to the data (a reduced $\chi^{2}$ of 1.07 for 684 degrees of freedom), and is shown in the bottom panel of Fig. \ref{abs_spec_fig_2}. The best fitting component values are very realistic: the partial covering model gives a column density of $2^{+0.4}_{-0.3}\times10^{22}$ cm$^{-2}$ (similar to the $\sim10^{22}$ cm$^{-2}$ we estimated earlier from the mass of molecular gas present) with a covering fraction of $0.15^{+0.02}_{-0.02}$. The temperature of the added APEC component is $1.0^{+0.05}_{-0.05}$ keV (with a metal abundance of 0.5 Z$_{\odot}$) which is reasonably similar to what is obtained by fitting a typical filament spectrum from Perseus with a single absorbed APEC component. The intensity of this component is 4.3$\times$10$^{-16}$ ergs cm$^{-2}$ s$^{-1}$ arcsec$^{-2}$ in the 0.6-2.0 keV band, which is very similar to the average intensity of the 12 filaments in Perseus we studied earlier, which is 4.0$\times$10$^{-16}$ ergs cm$^{-2}$ s$^{-1}$ arcsec$^{-2}$ in the same band.

We therefore conclude that the X-ray spectrum of the molecular region is consistent with absorption from molecular gas, which is partially compensated for by X-ray emission from the filaments.

\section{Conclusions}
\label{conclusion_section}

We have explored the X-ray properties of the filaments surrounding the BCGs in four nearby, X-ray bright galaxy clusters using deep Chandra observations: the Perseus cluster, the Centaurus cluster, the Virgo cluster and Abell 1795. In the Perseus cluster, we have studied the X-ray properties of 12 individual filaments, all of which can be fitted with an absorbed two component thermal model. The temperatures of these two components are consistent within errors between all of the 12 filaments, with a lower temperature of 0.77 keV, and an upper temperature 1.77 keV. This absorbed two component thermal model also proved to be a good fit to the spectra of the stacked filaments in Virgo, Centaurus and Abell 1795, and similar temperatures of the two components were found. The filaments in all of these systems therefore appear to have the same spectral shape, suggesting that the physical process causing the soft X-ray emission from the filaments is the same in all of these systems. 

We find that it is also possible to fit the stacked Chandra ACIS filament spectrum from Perseus with a model consisting of a charge exchange component and a soft thermal component. To obtain a good fit below 1 keV, the oxygen abundance needs to be suppressed relative to Fe, to reduce the intensity of the OVII and OVIII lines, an effect which cannot be achieved by just raising the level of absorption. Below 1 keV, it is possible to adequately fit the filament spectrum with just a variable abundance charge exchange model (bottom left hand panel of Fig. \ref{Filaments_stack_Perseus_CX}). However, to obtain the correct spectral shape above 1keV, an additional low temperature APEC component is needed (bottom right hand panel of Fig. \ref{Filaments_stack_Perseus_CX}). The charge exchange still provides the dominant contribution to the spectrum in the 0.5-1.0 keV band. Whilst the fit quality is not as good as an absorbed two temperature thermal model, further improvements to the charge exchange model in the future may improve the quality of the fit. These findings therefore add support to the idea that charge exchange at least contributes to the X-ray emission we see from the filaments.

We find that when we examine the high spectral resolution RGS spectrum of the filamentary plume seen in X-rays in the Centaurus cluster, the OVII line, which is expected to be prominent in the charge exchange spectrum, is not present. It therefore appears unlikely that the soft X-ray emission seen in the X-ray plume in Centaurus is dominated by charge exchange emission. This X-ray plume does not however contain any H-$\alpha$ filaments, making it fundamentally different to the filamentary regions of the Perseus cluster we explore, and possibly explaining the weakness of any charge exchange spectral features. The RGS spectrum also contains a significant background contribution from the surrounding ICM which cannot be accurately removed, and may prevent us from detecting the OVII line.

In the Perseus cluster, we search for signs of X-ray absorption due to gas in the filaments, using a region which has a large amount of molecular gas, and comparing it to a nearby offset region where there are no filaments. At energies below 2 keV, there is a deficit in the X-ray spectrum of the molecular gas region, suggesting that absorption is playing a role. However the X-ray spectrum of this deficit is complex, and cannot be fit by just simply adding additional absorbing components. 

This suggests that, in addition to absorption, the gas also emits X-rays which partially compensates for the absorption, leading to a more complex spectral shape. We find that when we add an additional soft thermal component to represent X-ray filament emission, we obtain a good fit to the X-ray spectrum of the molecular gas region. Moreover, the physical parameters of the best fit model are very realistic. The column density of the absorption model agrees well with what would be expected given the molecular gas mas in the region obtained by \citet{Salome2008}. The X-ray surface brightness of the filament component is in good agreement with the average X-ray surface brightness of the 12 filaments we have studied in Perseus.

\section*{Acknowledgements}

SAW and ACF acknowledge support from ERC Advanced Grant
FEEDBACK. This work is based on observations obtained with the
Chandra observatory, a NASA mission. It is also based on observations with XMM-Newton, an ESA science mission with instruments and contributions
directly funded by ESA Member States and NASA.

\bibliographystyle{mn2e}
\bibliography{Filament_paper}

\appendix

\section[]{}

\label{sec:appendix}

\begin{table*}
\begin{center}
\caption{Chandra data used in this paper.}
\label{All_obsids}
\leavevmode
\begin{tabular}{l| l|l | l | l | l} \hline \hline
Object & Obs ID&Exposure (ks)&RA&Dec&Start Date\\ \hline
Perseus & 3209&95.77&03 19 47.60&+41 30 37.00&2002-08-08\\
& 4289&95.41&03 19 47.60&+41 30 37.00&2002-08-10\\
& 4946&23.66&03 19 48.20&+41 30 42.20&2004-10-06\\
& 4947&29.79&03 19 48.20&+41 30 42.20&2004-10-11\\
& 6139&56.43&03 19 48.20&+41 30 42.20&2004-10-04\\
& 6145&85.00&03 19 48.20&+41 30 42.20&2004-10-19\\
& 4948&118.61&03 19 48.20&+41 30 42.20&2004-10-09\\ 
& 4949&29.38&03 19 48.20&+41 30 42.20&2004-10-12\\
& 6146&47.13&03 19 48.20&+41 30 42.20&2004-10-20\\
& 4950&96.92&03 19 48.20&+41 30 42.20&2004-10-12\\
& 4951&96.12&03 19 48.20&+41 30 42.20&2004-10-17\\
& 4952&164.24&03 19 48.20&+41 30 42.20&2004-10-14\\
& 4953&30.08&03 19 48.20&+41 30 42.20&2004-10-18\\ \hline
Centaurus & 504&31.75&12 48 48.70&-41 18 44.00&2000-05-22\\
& 4954&89.05&12 48 48.90&-41 18 44.40&2004-04-01\\
& 4955&44.68&12 48 48.90&-41 18 44.40&2004-04-02\\
& 5310&49.33&12 48 48.90&-41 18 44.40&2004-04-04\\
& 16223 &	180.0 	&12 48 48.90 &	-41 18 43.80 &	2014-05-26 \\
& 16224 &	 	42.29 &	12 48 48.90 	&-41 18 43.80 	& 	2014-04-09 \\ 	
& 16225 &	 	30.1 	&12 48 48.90 &	-41 18 43.80 	& 	 	2014-04-26 \\
& 16534 &	 	55.44 &	12 48 48.90 &	-41 18 43.80 	 &	 	2014-06-05 \\
& 16607 &	 	45.67 &	12 48 48.90 &	-41 18 43.80 	 &	 	2014-04-12 \\
& 16608 &	 	34.11 &	12 48 48.90 &	-41 18 43.80 	 &	 	2014-04-07 \\
& 16609 &	 	82.33 &	12 48 48.90 &	-41 18 43.80 	 &	 	2014-05-04 \\ 	
& 16610 &	 	17.34 &	12 48 48.90 &	-41 18 43.80 	 &	 	2014-04-27\\ \hline
Virgo & 5826&126.76&12 30 49.50&+12 23 28.00&2005-03-03\\ 
& 5827&156.20&12 30 49.50&+12 23 28.00&2005-05-05\\
& 5828&32.99&12 30 49.50&+12 23 28.00&2005-11-17\\
& 6186&51.55&12 30 49.50&+12 23 28.00&2005-01-31\\
& 7210&30.71&12 30 49.50&+12 23 28.00&2005-11-16\\
& 7211&16.62&12 30 49.50&+12 23 28.00&2005-11-16\\
& 7212&65.25&12 30 49.50&+12 23 28.00&2005-11-14\\ \hline
Abell 1795 & See tables A1 and A2 from \citet{Walker2014_A1795} &&&&\\ \hline
Coma & 555&8.66&12 59 48.00&+27 58 00.00&1999-11-04\\
& 1112&9.65&12 59 48.00&+27 58 00.00&1999-11-04\\
& 1113&9.65&12 59 48.00&+27 58 00.00&1999-11-04\\
& 1114&9.05&12 59 48.00&+27 58 00.00&1999-11-04\\
& 9714&29.65&12 59 48.00&+27 58 00.00&2008-03-20\\
& 13993&39.56&12 59 51.92&+27 54 17.76&2012-03-21\\
& 13994&81.99&12 59 39.42&+27 54 28.59&2012-03-19\\
& 13995&62.98&12 59 59.57&+27 55 02.47&2012-03-14\\
& 13996&123.06&12 59 51.01&+27 56 33.68&2012-03-27\\
& 14410&78.53&12 59 51.92&+27 54 17.76&2012-03-22\\
& 14411&33.64&12 59 39.42&+27 54 28.59&2012-03-20\\
& 14415&34.53&12 59 59.57&+27 55 02.47&2012-04-13\\
& 14406&24.76&12 59 59.57&+27 55 02.47&2012-03-15\\ \hline
Abell 3627 & 8178&57.41&16 15 03.80&-60 54 25.20&2007-07-08\\
& 9518&140.04&16 13 25.59&-60 45 43.10&2008-06-13\\
& 12950&89.85&16 13 39.28&-60 52 10.20&2011-01-10\\ \hline
\end{tabular}
\end{center}
\end{table*}

\end{document}






















%% file: Filament_paper.bbl
\begin{thebibliography}{}

\bibitem[\protect\citeauthoryear{{Conselice}, {Gallagher} III \&
  {Wyse}}{{Conselice} et~al.}{2001}]{Conselice2001}
{Conselice} C.~J.,  {Gallagher} III J.~S.,    {Wyse} R.~F.~G.,  2001, \aj, 122,
  2281

\bibitem[\protect\citeauthoryear{{Crawford}, {Sanders} \& {Fabian}}{{Crawford}
  et~al.}{2005}]{Crawford2005}
{Crawford} C.~S.,  {Sanders} J.~S.,    {Fabian} A.~C.,  2005, \mnras, 361, 17

\bibitem[\protect\citeauthoryear{{Dennerl}, {Englhauser} \&
  {Tr{\"u}mper}}{{Dennerl} et~al.}{1997}]{Dennerl1997}
{Dennerl} K.,  {Englhauser} J.,    {Tr{\"u}mper} J.,  1997, Science, 277, 1625

\bibitem[\protect\citeauthoryear{{Donahue}, {Connor}, {Fogarty}, {Li}, {Voit},
  {Postman}, {Koekemoer}, {Moustakas}, {Bradley} \& {Ford}}{{Donahue}
  et~al.}{2015}]{Donahue2015}
{Donahue} M.,  {Connor} T.,  {Fogarty} K.,  {Li} Y.,  {Voit} G.~M.,  {Postman}
  M.,  {Koekemoer} A.,  {Moustakas} J.,  {Bradley} L.,    {Ford} H.,  2015,
  \apj, 805, 177

\bibitem[\protect\citeauthoryear{{Fabian}, {Johnstone}, {Sanders}, {Conselice},
  {Crawford}, {Gallagher} III \& {Zweibel}}{{Fabian} et~al.}{2008}]{Fabian2008}
{Fabian} A.~C.,  {Johnstone} R.~M.,  {Sanders} J.~S.,  {Conselice} C.~J.,
  {Crawford} C.~S.,  {Gallagher} III J.~S.,    {Zweibel} E.,  2008, \nat, 454,
  968

\bibitem[\protect\citeauthoryear{{Fabian}, {Sanders}, {Crawford}, {Conselice},
  {Gallagher} \& {Wyse}}{{Fabian} et~al.}{2003}]{Fabian2003}
{Fabian} A.~C.,  {Sanders} J.~S.,  {Crawford} C.~S.,  {Conselice} C.~J.,
  {Gallagher} J.~S.,    {Wyse} R.~F.~G.,  2003, \mnras, 344, L48

\bibitem[\protect\citeauthoryear{{Fabian}, {Sanders}, {Williams}, {Lazarian},
  {Ferland} \& {Johnstone}}{{Fabian} et~al.}{2011}]{Fabian2011_filaments}
{Fabian} A.~C.,  {Sanders} J.~S.,  {Williams} R.~J.~R.,  {Lazarian} A.,
  {Ferland} G.~J.,    {Johnstone} R.~M.,  2011, \mnras, 417, 172

\bibitem[\protect\citeauthoryear{{Ferland}, {Fabian}, {Hatch}, {Johnstone},
  {Porter}, {van Hoof} \& {Williams}}{{Ferland} et~al.}{2008}]{Ferland2008}
{Ferland} G.~J.,  {Fabian} A.~C.,  {Hatch} N.~A.,  {Johnstone} R.~M.,  {Porter}
  R.~L.,  {van Hoof} P.~A.~M.,    {Williams} R.~J.~R.,  2008, \mnras, 386, L72

\bibitem[\protect\citeauthoryear{{Ferland}, {Fabian}, {Hatch}, {Johnstone},
  {Porter}, {van Hoof} \& {Williams}}{{Ferland} et~al.}{2009}]{Ferland2009}
{Ferland} G.~J.,  {Fabian} A.~C.,  {Hatch} N.~A.,  {Johnstone} R.~M.,  {Porter}
  R.~L.,  {van Hoof} P.~A.~M.,    {Williams} R.~J.~R.,  2009, \mnras, 392, 1475

\bibitem[\protect\citeauthoryear{{Hatch}, {Crawford}, {Johnstone} \&
  {Fabian}}{{Hatch} et~al.}{2006}]{Hatch2006}
{Hatch} N.~A.,  {Crawford} C.~S.,  {Johnstone} R.~M.,    {Fabian} A.~C.,  2006,
  \mnras, 367, 433

\bibitem[\protect\citeauthoryear{{Johnstone}, {Hatch}, {Ferland}, {Fabian},
  {Crawford} \& {Wilman}}{{Johnstone} et~al.}{2007}]{Johnstone2007}
{Johnstone} R.~M.,  {Hatch} N.~A.,  {Ferland} G.~J.,  {Fabian} A.~C.,
  {Crawford} C.~S.,    {Wilman} R.~J.,  2007, \mnras, 382, 1246

\bibitem[\protect\citeauthoryear{{Kalberla}, {Burton}, {Hartmann}, {Arnal},
  {Bajaja}, {Morras} \& {P{\"o}ppel}}{{Kalberla} et~al.}{2005}]{LABsurvey}
{Kalberla} P.~M.~W.,  {Burton} W.~B.,  {Hartmann} D.,  {Arnal} E.~M.,  {Bajaja}
  E.,  {Morras} R.,    {P{\"o}ppel} W.~G.~L.,  2005, \aap, 440, 775

\bibitem[\protect\citeauthoryear{{Lazarian}, {Kowal}, {Vishniac} \& {de Gouveia
  Dal Pino}}{{Lazarian} et~al.}{2011}]{Lazarian2011}
{Lazarian} A.,  {Kowal} G.,  {Vishniac} E.,    {de Gouveia Dal Pino} E.,  2011,
   59, 537

\bibitem[\protect\citeauthoryear{{Lazarian}, {Santos-Lima} \& {de Gouveia Dal
  Pino}}{{Lazarian} et~al.}{2010}]{Lazarian2010}
{Lazarian} A.,  {Santos-Lima} R.,    {de Gouveia Dal Pino} E.,  2010, in
  {Pogorelov} N.~V.,  {Audit} E.,   {Zank} G.~P.,  eds, Numerical Modeling of
  Space Plasma Flows, Astronum-2009 Vol.~429 of Astronomical Society of the
  Pacific Conference Series, {Reconnection Diffusion and Star Formation
  Processes}.
p.~113

\bibitem[\protect\citeauthoryear{{McDonald}, {Veilleux}, {Rupke}, {Mushotzky}
  \& {Reynolds}}{{McDonald} et~al.}{2011}]{McDonald2011}
{McDonald} M.,  {Veilleux} S.,  {Rupke} D.~S.~N.,  {Mushotzky} R.,
  {Reynolds} C.,  2011, \apj, 734, 95

\bibitem[\protect\citeauthoryear{{McNamara}, {Wise} \& {Murray}}{{McNamara}
  et~al.}{2004}]{McNamara2004}
{McNamara} B.~R.,  {Wise} M.~W.,    {Murray} S.~S.,  2004, \apj, 601, 173

\bibitem[\protect\citeauthoryear{{O'Dea}, {Baum}, {Mack}, {Koekemoer} \&
  {Laor}}{{O'Dea} et~al.}{2004}]{ODea2004}
{O'Dea} C.~P.,  {Baum} S.~A.,  {Mack} J.,  {Koekemoer} A.~M.,    {Laor} A.,
  2004, \apj, 612, 131

\bibitem[\protect\citeauthoryear{{O'Dea}, {Quillen}, {O'Dea}, {Tremblay},
  {Snios}, {Baum}, {Christiansen}, {Noel-Storr}, {Edge}, {Donahue} \&
  {Voit}}{{O'Dea} et~al.}{2010}]{ODea2010}
{O'Dea} K.~P.,  {Quillen} A.~C.,  {O'Dea} C.~P.,  {Tremblay} G.~R.,  {Snios}
  B.~T.,  {Baum} S.~A.,  {Christiansen} K.,  {Noel-Storr} J.,  {Edge} A.~C.,
  {Donahue} M.,    {Voit} G.~M.,  2010, \apj, 719, 1619

\bibitem[\protect\citeauthoryear{{Salom{\'e}}, {Revaz}, {Combes}, {Pety},
  {Downes}, {Edge} \& {Fabian}}{{Salom{\'e}} et~al.}{2008}]{Salome2008}
{Salom{\'e}} P.,  {Revaz} Y.,  {Combes} F.,  {Pety} J.,  {Downes} D.,  {Edge}
  A.~C.,    {Fabian} A.~C.,  2008, \aap, 483, 793

\bibitem[\protect\citeauthoryear{{Sanders}, {Fabian}, {Allen}, {Morris},
  {Graham} \& {Johnstone}}{{Sanders} et~al.}{2008}]{Sanders2008}
{Sanders} J.~S.,  {Fabian} A.~C.,  {Allen} S.~W.,  {Morris} R.~G.,  {Graham}
  J.,    {Johnstone} R.~M.,  2008, \mnras, 385, 1186

\bibitem[\protect\citeauthoryear{{Sanders}, {Fabian}, {Sun}, {Churazov},
  {Simionescu}, {Walker} \& {Werner}}{{Sanders} et~al.}{2014}]{Sanders2014}
{Sanders} J.~S.,  {Fabian} A.~C.,  {Sun} M.,  {Churazov} E.,  {Simionescu} A.,
  {Walker} S.~A.,    {Werner} N.,  2014, \mnras, 439, 1182

\bibitem[\protect\citeauthoryear{{Smith}, {Foster} \& {Brickhouse}}{{Smith}
  et~al.}{2012}]{ACXpaper}
{Smith} R.~K.,  {Foster} A.~R.,    {Brickhouse} N.~S.,  2012, Astronomische
  Nachrichten, 333, 301

\bibitem[\protect\citeauthoryear{{Sun}, {Donahue}, {Roediger}, {Nulsen},
  {Voit}, {Sarazin}, {Forman} \& {Jones}}{{Sun} et~al.}{2010}]{SunA3627}
{Sun} M.,  {Donahue} M.,  {Roediger} E.,  {Nulsen} P.~E.~J.,  {Voit} G.~M.,
  {Sarazin} C.,  {Forman} W.,    {Jones} C.,  2010, \apj, 708, 946

\bibitem[\protect\citeauthoryear{{Tremblay}, {O'Dea}, {Baum}, {Mittal},
  {McDonald}, {Combes}, {Li}, {McNamara}, {Bremer}, {Clarke}, {Donahue},
  {Edge}, {Fabian}, {Hamer}, {Hogan}, {Oonk}, {Quillen}, {Sanders},
  {Salom{\'e}} \& {Voit}}{{Tremblay} et~al.}{2015}]{Tremblay2015}
{Tremblay} G.~R.,  {O'Dea} C.~P.,  {Baum} S.~A.,  {Mittal} R.,  {McDonald}
  M.~A.,  {Combes} F.,  {Li} Y.,  {McNamara} B.~R.,  {Bremer} M.~N.,  {Clarke}
  T.~E.,  {Donahue} M.,  {Edge} A.~C.,  {Fabian} A.~C.,  {Hamer} S.~L.,
  {Hogan} M.~T.,  {Oonk} J.~B.~R.,  {Quillen} A.~C.,  {Sanders} J.~S.,
  {Salom{\'e}} P.,    {Voit} G.~M.,  2015, \mnras, 451, 3768

\bibitem[\protect\citeauthoryear{{Walker}, {Fabian} \& {Kosec}}{{Walker}
  et~al.}{2014}]{Walker2014_A1795}
{Walker} S.~A.,  {Fabian} A.~C.,    {Kosec} P.,  2014, \mnras, 445, 3444

\bibitem[\protect\citeauthoryear{{Werner}, {Oonk}, {Canning}, {Allen} \&
  {Simionescu}}{{Werner} et~al.}{2013}]{Werner2013}
{Werner} N.,  {Oonk} J.~B.~R.,  {Canning} R.~E.~A.,  {Allen} S.~W.,
  {Simionescu} A.,  2013, \apj, 767, 153

\bibitem[\protect\citeauthoryear{{Zhang}, {Sun}, {Ji}, {Sarazin}, {Lin},
  {Nulsen}, {Roediger}, {Donahue}, {Forman}, {Jones}, {Voit} \& {Kong}}{{Zhang}
  et~al.}{2013}]{ZhangA3627}
{Zhang} B.,  {Sun} M.,  {Ji} L.,  {Sarazin} C.,  {Lin} X.~B.,  {Nulsen}
  P.~E.~J.,  {Roediger} E.,  {Donahue} M.,  {Forman} W.,  {Jones} C.,  {Voit}
  G.~M.,    {Kong} X.,  2013, \apj, 777, 122

\end{thebibliography}
